# Estimation of hysteretic losses in the HTS coils made of coated conductor tapes of an electric generator during transient operation

Víctor M. R. Zermeño[1*], Asger B. Abrahamsen[2], Nenad Mijatovic[3], Bogi B. Jensen[4] and Mads P. Sørensen[5]

[1] Karlsruhe Institute of Technology, P.O. Box 3640, 76021 Karlsruhe, Germany
[2] Department of Wind Energy, Technical University of Denmark, Frederiksborgvej 399, 4000 Roskilde, Denmark
[3] Department of Electrical Engineering, Technical University of Denmark, Ørsteds Plads, Bldg. 348, 2800 Kgs. Lyngby, Denmark
[4] Department of Science and Technology, University of the Faroe Islands, FO-100 Torshavn, The Faroe Islands
[5] Department of Applied Mathematics and Computer Science, Technical University of Denmark, Richard Petersens Plads, Bldg. 324, 2800 Kgs. Lyngby, Denmark

Corresponding author: victor.zermeno@kit.edu

**Abstract**

In this work we present a modeling tool designed to estimate the hysteretic losses in the coils of an electric generator with coils made of coated conductor tapes during transient operation. The model is based on a two-stage segregated model approach that allows simulating the electric generator and the current distribution in the superconducting coils using a one-way coupling from the generator to the HTS coils model. The model has two inputs: the rotational speed and the electric load signal. A homogeneous anisotropic bulk model for the coils allows computing the current distribution in the coils. From this distribution, the hysteretic losses are estimated. Beyond the interest on providing an estimate on the global energy dissipation in the machine, in this work we present a more detailed local analysis that allows addressing issues such as coil design, critical current ratting, electric load change rate limits, cryocooler design, identification of quench-prone regions and overall transient performance.

## 1. Introduction

Being lighter and more compact, and with higher efficiency at partial loads, superconducting generators could be advantageous with respect to their Cu based counterparts [1]–[4]. Nowadays there is growing interest in their use for several applications where size and weight are to be reduced. These applications include among others: generators for airplanes [5], ships [2],[6] and wind turbines [7],[8],[9]. To avoid AC losses, these machines are designed for operation under DC conditions. However, during normal operation, superconducting generators can be exposed to ripple fields produced from a wide variety of sources: the teeth in the stator, load change, rotational speed fluctuations, etc. Unlike in the DC conditions case, superconductors have high losses when exposed to AC fields. Thus, calculation of such losses is relevant for machine design, cryostat optimization and for increasing the overall performance.

Modeling and simulation of these transients is a challenging task. In the particular case of coated conductor tapes, it requires considering a system that spans spatially 5 (or 6) orders of magnitude: from the 1 μm thick superconducting layers in the windings, to the actual generators in the kW (MW) class with an expected cross section in the order of decimeters (meters).

Previous works aimed at estimating the losses in the superconducting coils of generators include among others [10] and [11]. These works dealt with estimating the hysteretic losses in the $MgB_2$ wires of the electric generator coils. In [10], an analytical method based on the Bean model was used. In [11], a detailed numerical simulation using a multiscale technique was implemented to provide an estimate. Said multiscale technique had been previously presented in [12] where it was used to estimate the losses in coils made of coated conductor tapes. It is also important to note a previous implementation of ours [13]. There, we proposed a two-stage segregated model approach that separated the electromagnetic fields calculation in the generator from the superconducting coils. There, a constant critical current density $J_c$ was assumed for the HTS layer of the coated conductor tape. Among other improvements, the present work now considers a realistic dependence of the critical current on the amplitude and direction of the local magnetic field $B$ as described in [14].

A recent comparison between the multiscale and homogenization methods [15] showed that although the multiscale method allows for a full parallelization of the computations and with it, a full scalability to large scale designs, the homogenization provided a better estimation of the ac losses in arrays of tapes, both locally and globally. Therefore, in the present work we have chosen to follow the homogenization method.

The main purpose of this study is to present a tool for estimating the hysteretic energy losses in the HTS coils made of coated conductor tapes of an electric generator during transient operation such as load changes during the operation. The model is driven by both rotational speed and electric load signals. A two-stage segregated model approach is used to separate the electromagnetic fields calculation in the generator from the superconducting coils. Calculation of AC losses is accomplished by using a homogeneous anisotropic bulk model.

Although general interest is given to estimate the global energy losses, particular emphasis is also made on the local effects of the losses. Understanding these allows addressing several important design and performance issues such as critical current of the superconducting coils, electric load change rate, cryostat design and identification of quench-prone regions.

## 2. Modeling strategy

The simulation goal was set on computing the AC losses in an electric generator with rotor windings made of coated conductor tapes during transient behavior. In that way, heating losses are to be computed as a response to the electric load and rotational speed signals. To cope with the size of the modeling problem at hand and for simplicity, a two-stage segregated model was pursued. In that way, the generator electromagnetic computation could be decoupled from the superconducting coils.

Considering tightly packed coils, if each conductor in the rotor windings carries a net given current $I(t)$, the magnetic field $H$ in the coupling boundary, as shown in Figure 1, has a weak dependence upon the actual current density distribution inside each conductor. Hence, the magnetic field $H$ at the coupling boundary is mainly due to the global effect of the current in the coils, rather than the actual current distribution inside each conductor. This allows simulating the electromagnetic response of the generator without taking into account the geometrical internal features of the rotor's coils, or the $E-J$ characteristic relationship of its constituent materials. In fact, that is the standard method used to simulate conventional rotating machinery, where a uniform current density is typically assumed in the rotor windings. Although some studies such as coil optimization in HTS based machines can be performed following this principle [16], calculation of AC losses in the superconducting materials requires knowledge of the local current density.

Following from the assumption described above, the actual current density distribution in the conductors of the rotor's coils can be recalculated from the generator simulation data. Using the previously computed magnetic field $H$ in the coupling boundary as a Dirichlet condition, a second simulation can be performed to calculate the AC losses. Consistent with the original assumption, no feedback is considered from the second to the first model. In what follows and for convenience, the first model will be referred as the "generator model" and the second as "the coils model". Hence, the overall strategy as depicted in Figure 1 can be summarized as follows:

1) Both rotational speed and electric load are used as input signals for the generator model where the current distribution in the rotor windings is assumed uniform.
2) The coils model uses the previously computed magnetic field $H$ on a coupling boundary to calculate the actual current density distribution and the AC losses in the superconducting windings of the rotor's coils.

For simplicity, in what follows, the coils will be referred to by their number as shown on the right side of Figure 1.

**Generator model**

The generator model (Figure 1) follows design parameters as reported in [17] and [18]. This provided with a test bench for running transient problems. The model is driven solely by the rotor speed and electric resistive load signals. For uniformity, every superconducting coil was assumed to be comprised of 50 turns of the coated conductor tapes manufactured by Superpower [2] each carrying 50 A during steady-state operation. In the generator model, this corresponded to a uniform current density such that a net 2500 Ampturns are carried in each of the four coils. The model uses the AC/DC module of COMSOL Multiphysics for rotating machinery. This module solves the magnetic vector potential formulation of Maxwell's equations [19]. The magnetic field $H$ is then calculated with $H = (\nabla \times A)/\mu$, where $A$ is the magnetic vector potential and $\mu$ is the permeability of the material. At the coupling boundary, data is stored for further analysis in the coils model.

**Coils model**

In principle, the coils could be modeled considering every layer of the coated conductor tape as described in [20]. However, this would require a rather long computing time. Instead, for the sake of increased computational speed, we have opted to model the coils following a homogenization method previously introduced in [14]. Use of the Homogenized model implied that all geometric features in the coils were "washed out" and four homogeneous anisotropic bulks were used to model the generator windings. All parameters, including the superconducting material properties, are as described in [14]. This corresponded to coils made of 4 mm wide coated conductor tape with a DC critical current in zero external field of $I_c(B = 0, T = 77K) = 99.2\,A$. A power law was used to characterize the $E-J$ relationship in the superconducting material. Also as described in [14], a realistic dependence of the critical current $J_c(\boldsymbol{B})$ on the amplitude and direction of the magnetic field $\boldsymbol{B}$ is considered. Integral constrains as previously described in the same reference were implemented in each bulk. In this way, an equivalent sinusoidal net current per conductor of 50 A was imposed. To couple with the generator model, appropriate boundary conditions were used at the coupling boundary.

**Case study: Coils Ramp-up and load change**

To test the proposed modeling strategy, a simple load change was implemented as shown in Figure 2. The main purpose of this test case is to show that the model is capable of simulating conditions that are close enough to real operation. Therefore, in this work we focus on the modeling capability rather than on discussing particular features of the simulated machine since most of them have already been discussed [18]. All events were simulated using smooth functions so that sharp transients could be avoided. Therefore, load changes were not instantaneous as small smoothing periods were allowed.

At $t = 0.0s$ the generator was stationary with its rotor coils already cooled down to 77 K. The coils were not energized and no electric load was connected to the terminals of the stator windings. From $t = 0.0s$ to $t = 1.0s$, the current in the rotor's coils was increased linearly until the desired value of 50 A was reached. At this point in time, the field winding provides magneto motive force of 10050 kAturns [18] which in turns results in 0.11Vrms/rpm phase value for beck electromotive force for each rpm of rotation. The rationale behind this selection for the current value will be explained in the discussion section. In what followed, that current was kept constant. From $t = 1.0s$ to $t = 7.0s$, the machine was kept static, with no electric load applied. This period was intended to let the system relax and allow for the transient behavior of the ramp up process to settle and fade. At $t = 7.0s$ the generator's shaft started rotating in counterclockwise direction at 1.91 Hz. At $t = 8.0s$, a resistive electric load $R_{load}$ of 1 Ω was connected to every phase of the generator (taking into account 0.32 Ω for the resistance of the copper windings). At t = 10.0s, the resistive load was changed to 0.5 Ω. Finally, at t = 12.0s, the load was reverted to its previous value of 1 Ω.

An equivalent circuit diagram of the generator and its load is presented in Figure 3. The transient load is represented here by the switches denoted by A and B. From t = 0 s to t = 8 s, both A and B are open. At t = 8 s A is closed, followed by B that is closed at t = 10 s and reopened at t = 12 s.

## 3. Results

Simulation of the whole transient of 14 seconds was carried out as described in the previous sections. In reporting the transient response, a selection of images describing it is presented in Figure 4. The images are presented in a sequential array, so that the transient can be followed in a comic-strip like fashion.

Figure 4 is divided horizontally in three sections: the magnetic flux density magnitude $|\boldsymbol{B}|$ as calculated by the coils model is shown at the top; the normalized critical current density $J/J_c(\boldsymbol{B})$ in the middle; and the local instantaneous power dissipation density ($\boldsymbol{E} \cdot \boldsymbol{J}$) at the bottom. Here, it is important to note that the color scale for this last plot is logarithmic. The images are arranged by columns so that they correspond to the same times (denoted at the top of each column). To better identify the overcritical regions, a white contour line is drawn at $J/J_c(\boldsymbol{B}) = 1$ in the images showing the normalized critical current density $J/J_c(\boldsymbol{B})$ (middle section). For compactness and to fit in the format of a journal article, only images related to the most relevant events are presented.

At $t = 0.5\ s$, the current in the coils is still ramping up. While the ramp-up of the coils takes place, transport currents appear in all coils. However, magnetization currents also appear in coils 1 and 4. For this reason, coils 1 and 4 show bigger overcritical regions than coils 2 and 3.

At $t = 7.8\ s$, the overcritical currents exhibited at $t = 1\ s$ have already relaxed and with them, the instantaneous energy dissipation has been reduced by almost three orders of magnitude. This is in agreement with the transient fading out. Also at $t = 7.8\ s$, when no load has been connected yet, no armature reaction is present and the magnetic flux density shows a symmetric profile.

Once the load is connected, At $t = 8.0\ s$, the magnetic flux density increases from the upper right corner of the coils. At this time overcritical current densities (dark red) are observed in coils 1 and 2, and to a smaller extent in the upper right corners of coils 3 and 4. In these overcritical regions the instantaneous energy dissipation is increased by almost three orders of magnitude. After this moment, small magnetization currents begin to appear in all coils and all subcritical regions are shrunk. By $t = 9.8\ s$, significant relaxation has occurred, allowing the instantaneous energy dissipation to be reduced again.

At $t = 10.0\ s$, when the load is increased, a higher magnitude magnetic flux density enters from the upper right corner. The overcritical zones grow covering the upper part of coil 1 and the upper right corners of the other coils. This turns into a substantial increase of the instantaneous energy dissipation, in particular in coils 1 and 2. Towards $t = 11.5\ s$, the overcritical current densities relax reducing the energy dissipation. When the load is reduced at $t = 12.0\ s$, small overcritical regions appear in the lower right corner of the coils yielding regions of increased dissipation. At $t = 14\ s$, the overcritical regions along with the corresponding energy dissipation fade after a small relaxation transient.

Overall, Figure 4 shows that during transient operation, regions with large normalized current density also show a large instantaneous power dissipation density: see for example the results at 1 s, 8 s, 10 s or 12 s. It is important to note that after the transient process has damped, said correlation mostly vanishes: see for example the results at 7.8 s, 9.8 s and to some extent 14 s.

A broader perspective can be provided by considering the instantaneous loss $\xi = \int \boldsymbol{E} \cdot \boldsymbol{J} dS$ in all coils and multiplying it by the length of the machine (15 cm). This corresponds to the total instantaneous loss in the straight section of the superconducting race track windings. Figure 5 presents such result. The highest loss was found during the ramp up of the coils at $t = 0.85$ s, reaching a peak value of 0.69 W. Peaks of 0.43 W ($t = 8.1\ s$) and 0.5 W ($t = 10.1\ s$) were also observed during the increasing load changes. A smaller peak of 0.016 W ($t = 12.16\ s$) was also observed during the decreasing load change. Qualitatively, all this is consistent in both schedule and amplitude with the planned transient and the results previously shown at the bottom section of Figure 4.

## 4. Discussion

During normal operation, but specially at increasing load changes, the upper part of coil 1 experienced the highest magnetic fields. Design tools such as coil optimization, addition of interpoles or the use of multiple excitation currents [16] could be used to allow the other coils to take a better share of this burden.

One important issue that can be clearly observed in the middle section of Figure 4, is that the coil design should consider the actual layout of the generator. The current density profile of an isolated coil as shown in Figure 6 of [14] has little to do with the actual profile it will display once it is located inside a generator. Hence, information of the DC critical current of an isolated coil is of little relevance without considering the electromagnetic environment it will finally face. Further improvements on coil design could be achieved by considering the normal operation requirements and the transitory load changes.

Several other values for the transport current in the superconducting coils were tested, however, the choice of 50 A was justified since during both ramp-up and increasing load change, large regions of the coils reached overcritical current density values as seen in the middle section of Figure 4 at $t = 8\ s$ and $t = 10\ s$. It could be argued that the overcritical zones can be avoided or reduced with longer and smoother transition periods. This idea is supported by noticing that after a small transitory, the current density in the coils returns to subcritical values, even at the highest load test. However, that is where the actual value of this simulation exercise tool pays off. Simulations can be performed to find the maximum load change rates allowed for a given current rating without running into overcritical currents.

Furthermore, this simulation tool can be used for cryocooling system design. Notice that during the ramp-up period, the hot zones show a symmetric pattern. The top of coils 1 and 2 and the bottom of coils 3 and 4 present the hottest regions. However, the hot areas in coils 3 and 4 vary for changing electric loads. This is especially noticeable in the upper right corner. Hence, cryocooler system integration could take this issue into account. For instance, heat sinks located on top of all coils and covering their right side could secure a better heat removal during transient response, than if placed only at the bottom of the coils. In the same manner, the hot regions are at risk of becoming normal if excess heat is not removed fast enough. Once a region becomes normal, ohmic losses appear with a corresponding further increase in temperature. This process can damage the superconducting materials; hence quench protection is essential for safe operation [21]. The simulation tool presented here could allow for *quench-prone* regions to be easily identified. Thus, sensors can be better located, and monitoring and quench protection optimized.

The last plot presented, Figure 5, can be used for comparison with experimental data, once it becomes available. It provides with a compact, lumped description of the overall hysteresis losses in a given machine design during transient operation.

Although in this work only the cases of ramp-up of rotor coils and electric load change were considered, a large collection of case scenarios can be analyzed using the same two-stage segregated model strategy, these include: short circuit of the stator's windings, connection to active loads, unbalanced loads, rotational speed change or the effect of including a damper cage, among others. Furthermore, when considering wind turbine applications, a mechanical model could be added to the generator model so that dynamics include the torque change due to wind speed variations.

A natural next step would be to estimate the hysteretic losses during transient operation of a larger scale device such as the multimegawatt superconducting wind turbine planned in the EU-funded EcoSwing project [9]. One important improvement to the model would be the coupling with a thermal transient model [22], [23]. This would allow a more precise design of the cryocooling system. A further natural next step should compare the computed transient losses with the Minimum Quench Energy of the coated conductor in the coil and the cooling capacity of the system. Finally, eventual estimation of the losses in the ends of the machine could be implemented using already available 3D models for racetrack coils [24] and a corresponding 3D model of the generator.

## 5. Conclusions

A two-stage segregated model strategy to calculate the hysteresis losses in a superconducting generator during transient operation was presented. The strategy relied upon a one-way coupling from the generator model to the coils model. The latter used a homogenization technique to model the coils in the rotor windings. The proposed model was used to simulate ramp-up and electric load change events. Analysis of results allowed considering important factors such as critical current ratting of a coil designed for operation in an electric generator. The issues of electric load change rate limit, cooling system design and identification of *quench-prone* regions was also addressed. Finally, calculation of the overall hysteresis losses in the HTS coils during transient response of the machine was presented. Further studies must be carried out to compare and match the simulation results with experimental data and calibrate the simulator to make tailored predictions for specific transient operation cases. As an additional remark, it is also important to notice that this strategy can easily be implemented to analyze and study other large scale superconducting machines and devices like motors, transformers, cables and magnet coils.

**Figures**

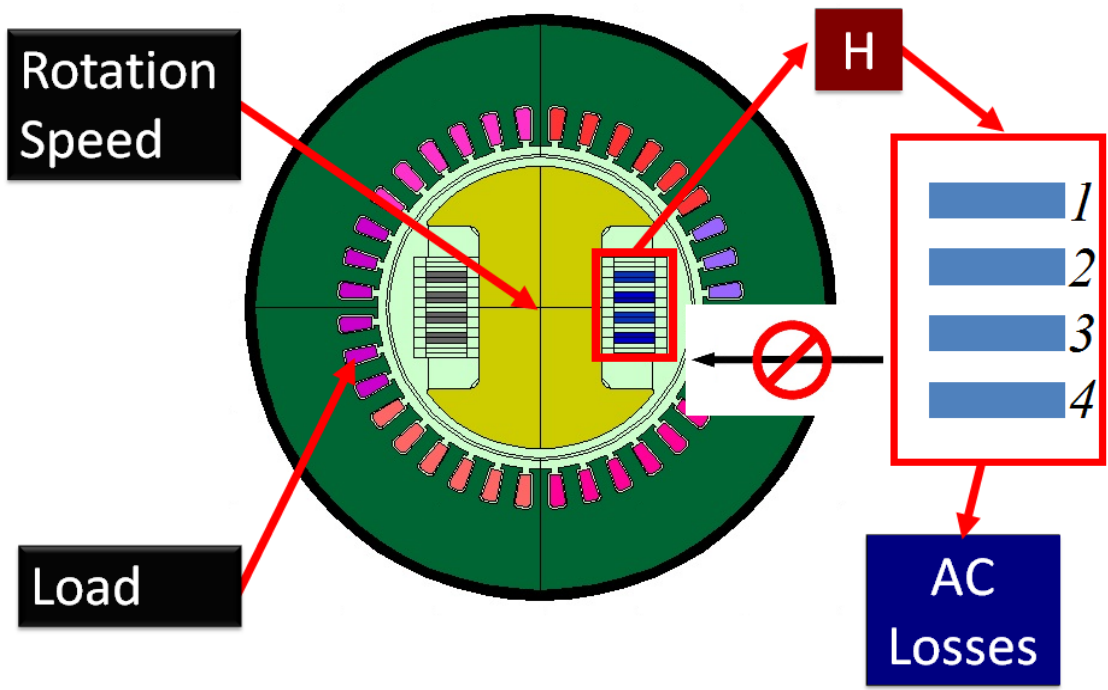

**Figure 1.** Simulation strategy. In the generator model (*center*), both electric load and rotation speed are considered as known inputs. Using this model, and assuming a uniform current distribution in the coils, the magnetic field *H* is calculated on a coupling boundary (*red rectangle*) around the rotor's windings. The coils model (*right*) uses this as a Dirichlet boundary condition to compute the actual current distribution in the rotor's coils and calculate the AC losses. No feedback is considered from the coils model to the generator model. The stator's outer and inner diameters are 28 cm and 16 cm respectively. The rotor's diameter is 14 cm. The axial length of the machine is 15 cm [18].

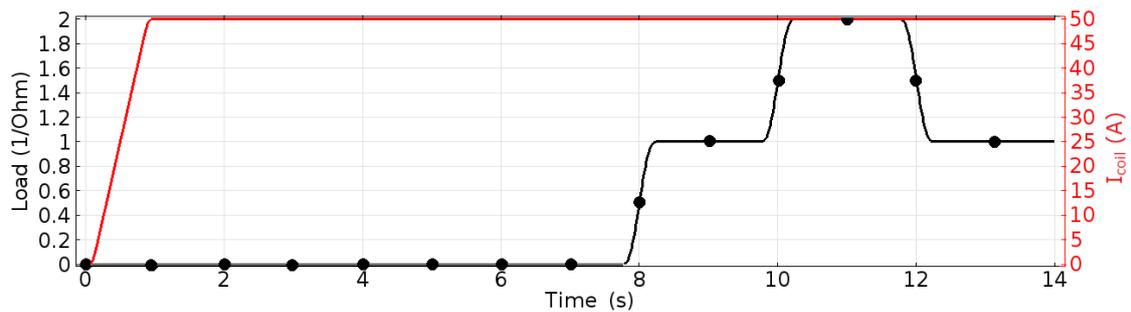

**Figure 2.** Current in each of the superconducting rotor coils (*red –*). The electric load signal in the stator windings (*black ●*) is presented as an equivalent conductance.

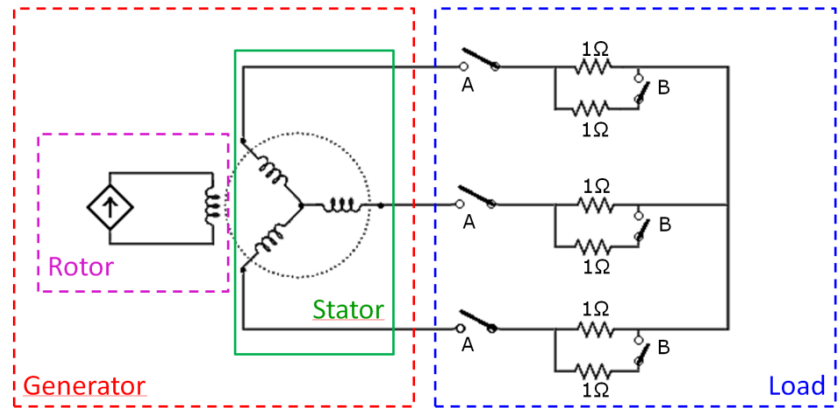

**Figure 3.** Equivalent circuit diagram representation of the electric generator and the applied resistive load.

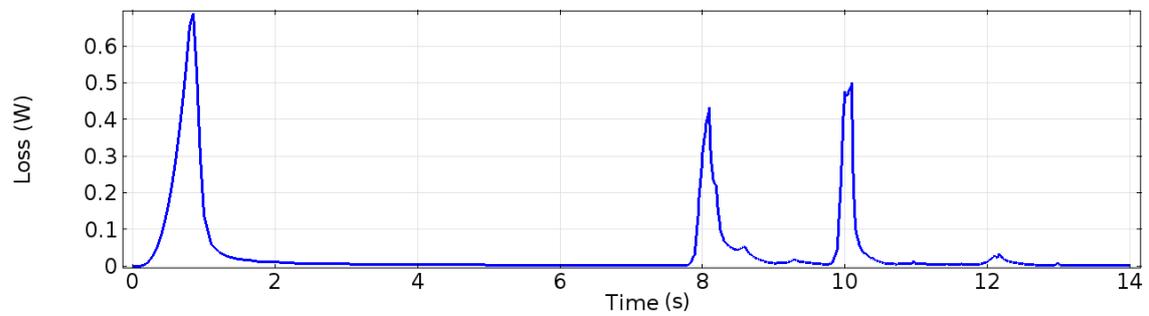

**Figure 5.** Total instantaneous loss in the superconducting coils of the generator.

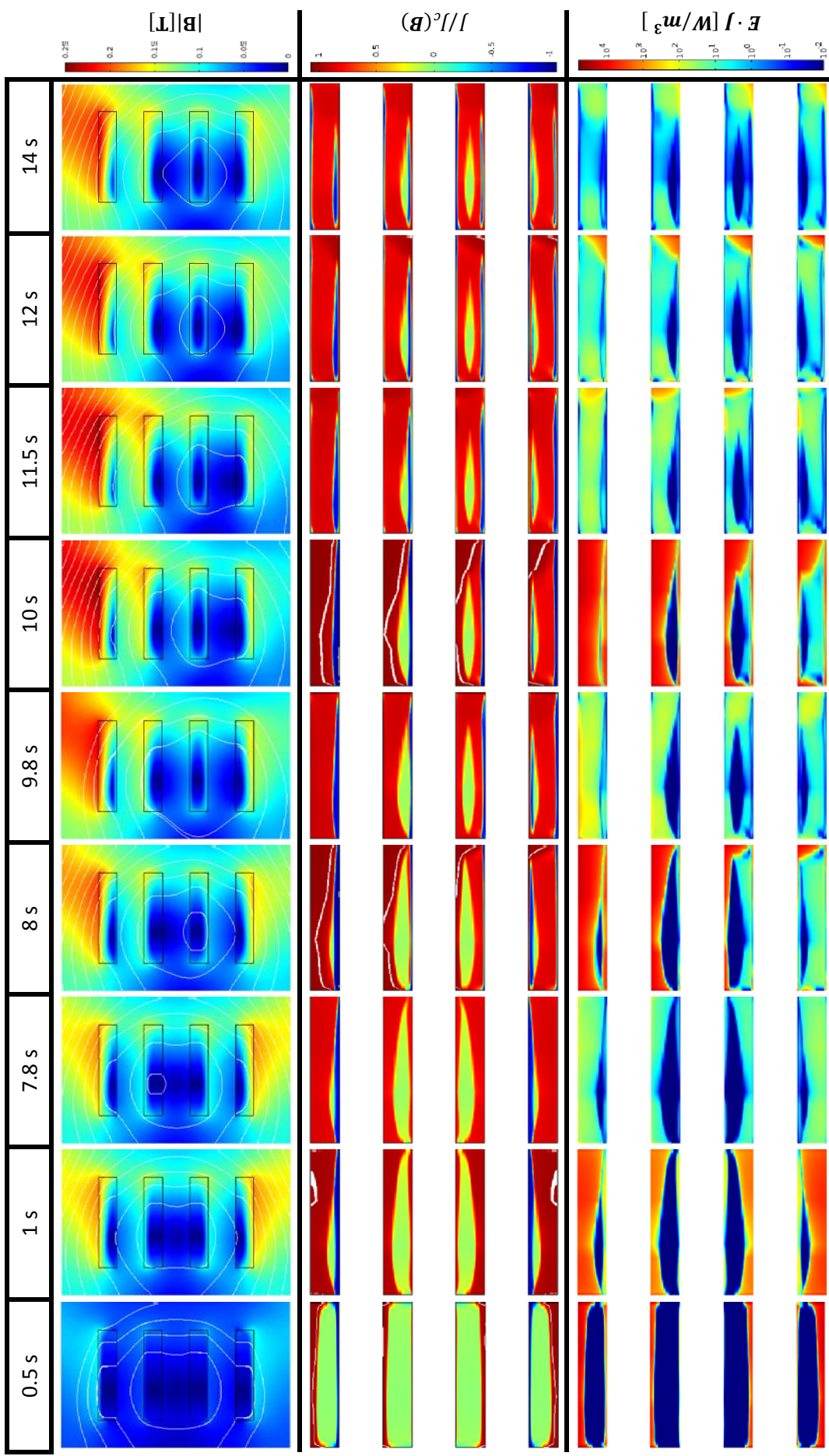

**Figure 4. Top:** Magnetic flux density magnitude $B$ [T] as calculated by the coils model. Streamlines follow the magnetic field lines. **Middle:** Normalized current density $J/J_c$ in the superconducting coils. As an aid to the eye to identify the supercritical regions, a white contour line is drawn at $J/J_c = 1$. **Bottom:** Instantaneous power dissipation density $(\boldsymbol{E} \cdot \boldsymbol{J})$ in the rotor's windings (in W/m$^3$). For easier visualization, a logaritmic scale is used. Data displayed was capped under the $10^{-2}$ value. Hence, all values from 0 to $10^{(-2)}$ were plotted with the darkest shade of blue.